\documentclass[aps,prl,twocolumn,showpacs,preprintnumbers,amsmath,amssymb,superscriptaddress]{revtex4}%

\usepackage{graphicx}%
\usepackage{dcolumn}
\usepackage{amsmath}
\usepackage{color}
\usepackage{multirow}

\makeatletter
\def\btt#1{\texttt{\@backslashchar#1}}%
\DeclareRobustCommand\bblash{\btt{\@backslashchar}}%
\makeatother

\begin{document}


\title{Tuning of competing magnetic and superconducting phase volumes in LaFeAsO$_{0.945}$F$_{0.055}$ by hydrostatic
pressure}

\author{R.~Khasanov}
 \email[Corresponding author: ]{rustem.khasanov@psi.ch}
 \affiliation{Laboratory for Muon Spin Spectroscopy, Paul Scherrer
Institute, CH-5232 Villigen PSI, Switzerland}
 \author{S.~Sanna}
  \affiliation{Dipartimento di Fisica ``A.\ Volta'' and Unit\`a CNISM di
 Pavia, I-27100 Pavia, Italy}
 \author{G.~Prando}
  \affiliation{Dipartimento di Fisica ``A.\ Volta'' and Unit\`a CNISM di
 Pavia, I-27100 Pavia, Italy}
  \affiliation{Dipartimento di Fisica ``E.\
 Amaldi'', Universit\`a di Roma3-CNISM, I-00146 Roma, Italy}
\author{Z.~Shermadini}
 \affiliation{Laboratory for Muon Spin Spectroscopy, Paul Scherrer
Institute, CH-5232 Villigen PSI, Switzerland}
\author{M.~Bendele}
 \affiliation{Laboratory for Muon Spin Spectroscopy, Paul Scherrer
Institute, CH-5232 Villigen PSI, Switzerland}
 \affiliation{Physik-Institut der Universit\"{a}t Z\"{u}rich,
Winterthurerstrasse 190, CH-8057 Z\"urich, Switzerland}
\author{A.~Amato}
 \affiliation{Laboratory for Muon Spin Spectroscopy, Paul Scherrer Institute, CH-5232 Villigen PSI, Switzerland}
\author{P.~Carretta}
 \affiliation{Dipartimento di Fisica ``A.\ Volta'' and Unit\`a CNISM di  Pavia, I-27100 Pavia, Italy}
 \author{R.~De Renzi}
  \affiliation{Dipartimento di Fisica and Unit\`a  CNISM di Parma,
 I-43124 Parma, Italy}
\author{J.~Karpinski}
 \affiliation{Laboratory for Solid State Physics, ETH Z\"urich, CH-8093 Z\"urich,
Switzerland}
\author{S.~Katrych}
 \affiliation{Laboratory for Solid State Physics, ETH Z\"urich, CH-8093 Z\"urich,
Switzerland}
\author{H.~Luetkens}
 \affiliation{Laboratory for Muon Spin Spectroscopy, Paul Scherrer Institute, CH-5232 Villigen PSI, Switzerland}
\author{N.D.~Zhigadlo}
 \affiliation{Laboratory for Solid State Physics, ETH Z\"urich, CH-8093 Z\"urich,
Switzerland}
\begin{abstract}
The interplay between magnetism and superconductivity  in
LaFeAsO$_{0.945}$F$_{0.055}$ was studied as a function of hydrostatic pressure
up to $p\simeq 2.4$~GPa by means of muon-spin rotation ($\mu$SR) and
magnetization measurements.
The application of pressure leads to a substantial decrease of the magnetic
ordering temperature $T_{\rm N}$ and a reduction of the magnetic phase volume
and, at the same time, to a strong increase of the superconducting transition
temperature $T_c$ and the diamagnetic susceptibility.
From the volume sensitive $\mu$SR measurements it can be concluded that the
superconducting and the magnetic areas which coexist in the same sample are
inclined towards spatial separation  and compete for phase volume as a function
of pressure.
\end{abstract}
\pacs{76.75.+i, 74.25.Ha, 74.62.Fj, 74.70.Xa }

\maketitle


The interplay between superconductivity and magnetism in high-temperature
superconductors (HTS) remains an important open issue. In cuprate and Fe-based
HTS the superconductivity can be induced in a magnetic parent compound by
charge doping and/or by pressure (chemical or external). In most cuprate HTS
the transformation from the magnetic into the superconducting state follows an
almost common scenario. On increasing the doping level the
antiferromagnetically ordered phase develops into a purely superconducting
state through a region where a spin-glass type of magnetism coexists with
superconductivity \cite{Niedermayer98,Coneri10,Khasanov_08-OIE-pahse-diagram}.
The situation with Fe-based HTS is, somehow, different. For some families of
Fe-based HTS like e.g. SmFeAsO$_{1-x}$F$_x$, Ba(Fe$_{1-x}$Co$_x$)$_2$As, and
FeSe$_{1-x}$Te$_x$, the magnetism is continuously suppressed and
superconductivity enhanced by changing the F, Co, or Se content. In the
intermediate region, bulk magnetism and bulk superconductivity are coexisting
in space \cite{Drew09,Sanna09,Nandi10,Khasanov09}. In
Ba$_{1-x}$K$_x$Fe$_2$As$_2$ the magnetic and the superconducting areas are
found to be separated microscopically as revealed, e.g., by atomic force
microscopy experiments \cite{Park09}.

One of the most interesting cases is realized in the LaFeAsO$_{1-x}$F$_x$
family of Fe-based HTS demonstrating an {\it abrupt} (first order like)
transition between the magnetic and the superconducting phases.
Muon-spin rotation ($\mu$SR) and M\"{o}ssbauer experiments show that above a
certain $x$ the samples become purely superconducting without visible traces of
magnetism \cite{Luetkens09}.
Such a behavior  seems to be rather different from the one observed for the
other structurally related families of Fe-based HTS in which the La atom is
replaced by Sm, Pr, Ce, etc. All of them demonstrate a coexistence between
superconductivity and magnetism for a certain doping level
\cite{Drew09,Sanna09,Sanna10}. Consequently the question if a similar
coexistence is present in LaFeAsO$_{1-x}$F$_x$ but within a much narrower, up
to now not detected, doping region or if an abrupt change between the
superconductivity and magnetism is a unique property of this particular family
of Fe-based HTS needs to be resolved.

Hydrostatic pressure experiments on LaFeAsO$_{0.945}$F$_{0.055}$, which is at
the border to the superconducting state but still magnetic, were performed to
distinguish between two above mentioned possibilities. This approach allows to
follow the transformation of the material from the magnetic to the
superconducting state in detail on one sample, i.e. without the necessity to
synthesize a large number of samples with exactly defined stoichiometry near
the phase boundary.
Our measurements show that both the magnetic and superconducting states are
most probably spatially separated in the crossover region of the phase diagram
and compete for phase volume.


The sample with the nominal composition LaFeAsO$_{0.945}$F$_{0.055}$ was
prepared in cubic anvil high-pressure cell from the stoichiometric mixture of
LaAs, FeAs, Fe$_2$O$_3$, Fe and LaF$_3$ \cite{Zhigadlo08}. A pressure of
$\simeq 3$~GPa was applied at  room temperature. By keeping the pressure
constant, the temperature was first ramped up to the maximum value of
1320$^{\rm o}$C, kept constant for 5.5~h and then quenched to room temperature
within a few minutes.


The superconducting properties of LaFeAsO$_{0.945}$F$_{0.055}$ were studied by
magnetization experiments. The zero-field-cooled and field-cooled (FC and ZFC)
DC magnetization measurements up to $p\simeq1.1$~GPa were performed by using
the commercial SQUID magnetometer (MPMS-XL7) and a piston-cylinder CuBe
pressure cell (''EasyLab Mcell 10``, \cite{EasyLab}). The AC experiments up to
$p\simeq2$~GPa were performed by using a home-made AC magnetometer (AC
frequency $\nu=72$~Hz, AC field amplitude $\mu_0H_{\rm AC}\simeq0.1$~mT). The
two pick-up and the excitation coils were wound directly around the pressure
cell made from MP35N alloy.
The sample and the small piece of In, used as the pressure
indicator, were located inside the different pick-up coils. Note that the AC
experiments within the present geometry (the coils wounded outside the pressure
cell) require separate measurements of the background signal from the empty
cell. For this particular sample, due to the very small superconducting
response at ambient pressure (see the discussion below), the AC magnetization
data measured at $p=0.0$~GPa were used as the background signal.

\begin{figure}[htb]
\includegraphics[width=1\linewidth]{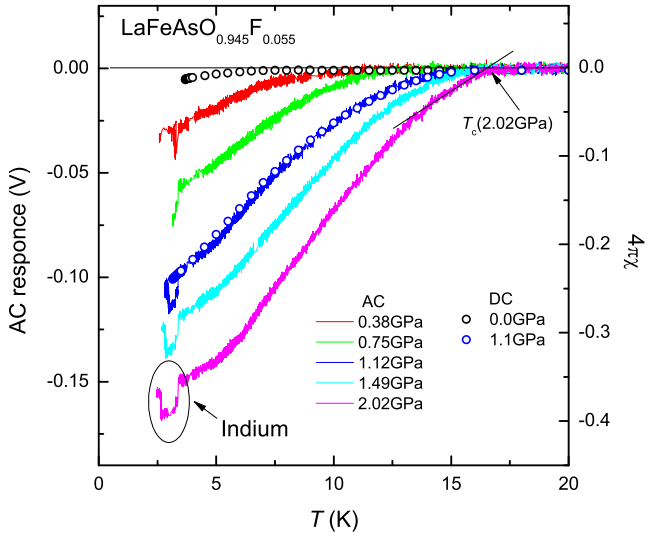}
%
\caption{(color online) AC (solid curves) and ZFC $\mu_0H=5$~mT DC (open
symbols) susceptibility curves as a function of temperature for different
pressures of LaFeAsO$_{0.945}$F$_{0.055}$. The transition temperature $T_c$ is
determined by the intersect of the linearly extrapolated magnetization curve
with the zero line. The superconducting transition of pure indium used as a
pressure indicator in the AC magnetization experiment is highlighted by the
oval for the highest pressure. }
 \label{fig:Magnetization}
\end{figure}

The results of the magnetization studies are presented in
Fig.~\ref{fig:Magnetization}. At ambient pressure the diamagnetic
susceptibility at $T\simeq3.5$~K [$\chi(3.5\ {\rm K})$] reaches approximately
1\% of its ideal value ($\chi_{\rm id}=-1/4\pi$). This suggests that the
superconductivity at $p=0.0$ is just filamentary and it is present only within
a small volume of the sample. With increasing pressure both, the onset
temperature of the superconducting transition  $T_c$ [determined from the
intersect of the linearly extrapolated $\chi(T)$ in the vicinity of $T_c$ with
the zero line, see Fig.~\ref{fig:Magnetization}] and the low-temperature value
of the diamagnetic response ($-4\pi\chi$), increase quite substantially.
According to Fig.~\ref{fig:Magnetization}, the increase of the external
pressure from $p=0.0$ to 2.02~GPa leads to
the shift of $T_c$ from $\simeq$7 to $\simeq$16~K
and to an increase of $-4\pi\chi$ from $\sim$1 to $\sim35$\%.
An additional set of DC magnetization measurements at $p\simeq 1.1$~GPa show
that the diamagnetic response reduces by a factor 2 for an applied field of
1~mT in ZFC and to a value of $4\pi\chi\simeq-0.05$ in $\mu_0H=5$~mT FC
experiments, respectively (both are not shown). Such differences, which could
be caused by the effect of pinning and the presence of  weak links between the
superconducting areas, do not permit a reliable evaluation of the genuine
superconducting volume. One may only conclude that the superconducting volume
fraction increases with increasing pressure but is always smaller than the
whole sample volume.


The magnetic properties of LaFeAsO$_{0.945}$F$_{0.055}$ were studied in
zero-field (ZF) muon-spin rotation experiments. Pressures up to $\simeq2.4$~GPa
were generated in a double wall piston-cylinder MP35N cell.
%
\begin{figure}[htb]
\includegraphics[width=0.95\linewidth]{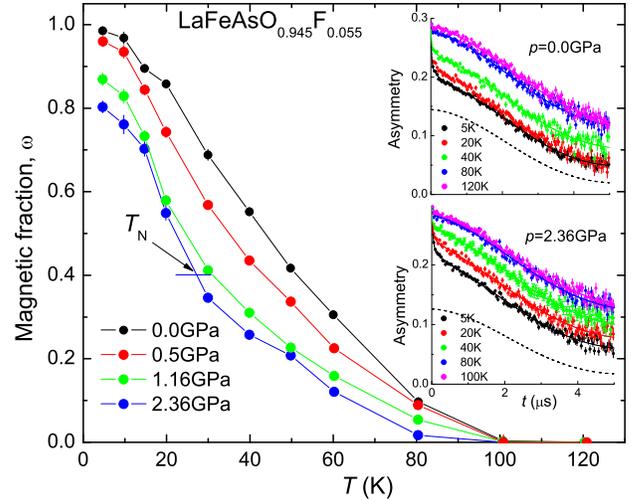}
%
\caption{(color online) Temperature dependence of the magnetic volume fraction
of LaFeAsO$_{0.945}$F$_{0.055}$ at various pressures. The insets show the ZF
muon-time spectra at $p=0.0$ and 2.36~GPa. The solid lines are fits by means of
Eq.~(\ref{eq:zf-musr}). The dotted lines represent the response of the pressure
cell.}
 \label{fig:MuSR}
\end{figure}
Few representative muon-time spectra measured at $p=0.0$ and 2.36~GPa are shown
in the insets of Fig.~\ref{fig:MuSR}. The data were analyzed by decomposing the
signal on the contribution of the sample and the pressure cell as:
\begin{equation}
 A(t)=A_{\rm S}(0)\;P_{\rm S}(t)+A_{\rm PC}(0)\;P_{\rm PC}(t).
 \label{eq:zf-musr}
\end{equation}
Here $A_{\rm S}(0)$ and $A_{\rm PC}(0)$ are the initial asymmetries and $P_{\rm
S}(t)$ and $P_{\rm PC}(t)$ are the muon-spin polarizations belonging to the
sample and the pressure cell, respectively. $P_{\rm PC}(t)$ was measured in an
independent experiment. The response of the sample was assumed to consist of a
magnetic and a nonmagnetic contribution and described as:
\begin{eqnarray}
 P_{\rm S}(t)&=&\omega\left[\frac{1}{3}\; e^{-\Lambda_{m,l}t}
 +\frac{2}{3}\; \biggl\{\zeta\;  e^{-\Lambda_{m,t1}t}+ (1-\zeta)\;
 e^{-\Lambda_{m,t2}t}\biggr\}\right]
 \nonumber \\
 &&+(1-\omega)\; e^{-\Lambda_{pm}t}. \nonumber
 \label{eq:zf-musr2}
\end{eqnarray}
Here $\omega$ is the relative weight (volume) of the magnetic fraction.
$\Lambda_{m,l}$, $\Lambda_{m,t1}$ and $\Lambda_{m,t2}$ are the exponential
depolarization rates representing the longitudinal (1/3) and the transversal
(2/3) relaxing components within the parts of the sample being in the magnetic
state. Two components within the curly brackets account for contributions of
two different muon stopping sites \cite{Maeter09} with the relative weight
$\zeta$ and ($1-\zeta$), respectively. $\Lambda_{\rm pm}$ is the relaxation
within the parts of the sample remaining nonmagnetic. The exponential character
of this relaxation instead of normally expected Kubo-Toyabe kind of behavior
\cite{Luetkens08} is probably caused by the presence of the small amount of
magnetic impurities, similar to the one observed in the so called '11' family
of Fe-based HTS (see e.g., Refs.~\onlinecite{Khasanov08,Khasanov09}).
For each particular pressure the whole set of the data was fitted
simultaneously with $A_{\rm S}(0)$, $A_{\rm PC}(0)$, $\zeta$ and the ratio
$\Lambda_{m,t1}/\Lambda_{m,t2}$  as common and $\omega$, $\Lambda_{m,l}$,
$\Lambda_{m,t1}$ and $\Lambda_{\rm pm}$ as individual parameters for each
temperature point. The solid lines in the insets of Fig.~\ref{fig:MuSR}
represent the result of the fit. The contribution from the cell at $T=5$~K is
shown as a dotted line.

The main panel of Fig.~\ref{fig:MuSR} shows the dependence of the magnetic
fraction $\omega$ on temperature for $p=0.0$, 0.5, 1.16, and 2.36~GPa.
Two important points needs to be considered. First of all, the magnetic volume
fraction at each particular temperature is lowered by the application of
pressure. Most noteworthy, with increasing pressure an increasingly large part
of the sample remains in the paramagnetic state down to lowest temperatures.
Second, the magnetic ordering temperature $T_{\rm N}$, defined as the
temperature where the magnetic fraction reaches 50\% of its maximum
low-temperature value, initially decreases with increasing pressure but then
demonstrates a tendency to saturate. $\omega(T)$ curves at $p=1.16$ 1.92 (not
shown) and 2.36~GPa being normalized to their values at $T\simeq5$~K become
almost identical.

\begin{figure}[htb]
\includegraphics[width=0.9\linewidth]{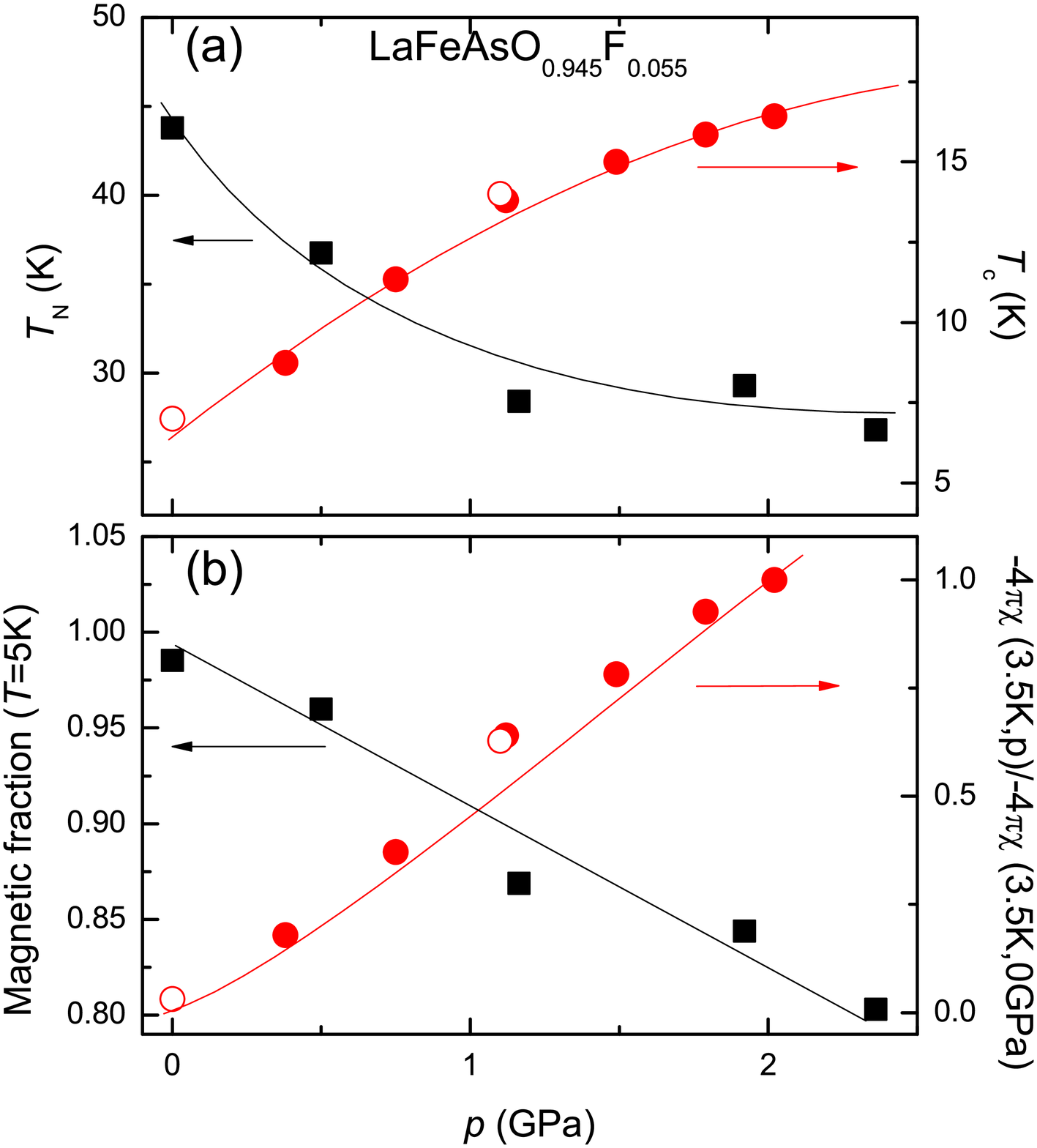}
%
\caption{(color online) (a) Dependence of the magnetic ordering temperature
$T_{\rm N}$ and the superconducting transition temperature $T_{\rm c}$ on
pressure. The closed and the open circles correspond to $T_{\rm c}$ as obtained
in AC and DC magnetization experiments, respectively. (b) The magnetic fraction
at $T=5$~K and ZFC diamagnetic susceptibility $-4\pi\chi$ at $T=3.5$~K,
$\mu_0H=5$~mT as a function of pressure. The closed and the open circles refer
to the data obtained in AC and DC magnetization experiments, respectively. The
lines are guide for the eye.}
 \label{fig:Correlations}
\end{figure}

In order to compare the influence of the pressure on the superconducting and
the magnetic properties of LaFeAsO$_{0.945}$F$_{0.055}$ the dependences of
$T_{\rm N}$, $T_{\rm c}$, $\omega$, and $4\pi\chi$ as a function of $p$ are
plotted in Fig.~\ref{fig:Correlations}. The decrease of $T_{\rm N}$ and
$\omega$ is associated with the corresponding increase of $T_{\rm c}$ and
$4\pi\chi$.
By applying a pressure of 2.36~GPa, $T_{\rm N}$ decreases from 44~K to 27~K,
while $T_{\rm c}$ more than doubles from $\simeq$7~K to $\simeq$16~K upon the
application of 2.02~GPa.

It should be noted here that the above presented data are pointing to a
competition of superconductivity and magnetism, but alone do not allow to
answer the question on how these two forms of order coexist within the
LaFeAsO$_{0.945}$F$_{0.055}$ sample.
There are three possible scenarios. The first one is the so called {\it phase
separation} scenario according to which the superconductivity develops just
within the parts of the sample remaining nonmagnetic down to low temperatures.
Such a  phase separated coexistence was observed, e.g.,  in
Ba$_{1-x}$K$_x$Fe$_2$As$_2$ \cite{Park09,Khasanov09_Ba122}.
The second possibility is an {\it atomic} coexistence of the superconducting
and magnetic order  parameters, which is consistent with models proposed in
Refs.~\onlinecite{Vorontsov09,Cvetkovic09} and most probably realized within
the so-called '11' family of Fe-based HTS \cite{Khasanov09,Bendele10}.
The third possibility is a {\it nanoscale} segregation into magnetic domains,
similar to that reported for cuprate HTS \cite{ Savici02,Coneri10,Russo07}. In
underdoped cuprate HTS, static, short-range, stripe-like magnetic correlations
are thought to exist in the superconducting state and are assumed not to affect
the superconducting carriers \cite{Coneri10}. Muons are sensitive to dipolar
fields at a distance of up to a few lattice spacings, so if nano-scale magnetic
domains exist then the fraction of muons experiencing static local magnetic
fields could be significantly higher than the fraction of Fe sites carrying an
ordered moment. Such type of coexistence was found to be realized within the
SmFeAsO$_{1-x}$F$_x$ and CeFeAsO$_{1-x}$F$_x$  families of Fe-based HTS
\cite{Sanna09,Sanna10}.

\begin{figure}[htb]
\includegraphics[width=1.0\linewidth]{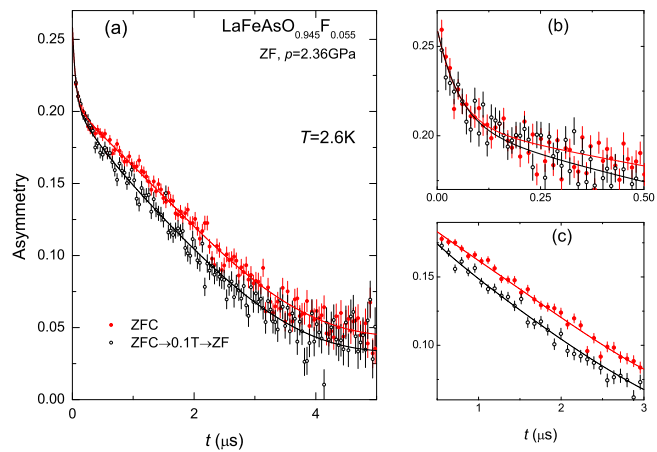}
%
\caption{(color online) (a) The ZF $\mu$SR time spectra obtained after cooling
the sample in zero magnetic field from $T\simeq100$~K down to 2.6~K (red
symbols) and after sweeping the magnetic field to 0.1~T and then setting it
back to 0.0 without changing the temperature (black symbols). The panels (b)
and (c) show the extended parts of the muon-time spectra at the lower and the
higher time, respectively. The solid lines are fits by means of
Eq.~(\ref{eq:zf-musr}).}
 \label{fig:ZF-history}
\end{figure}

Since the muon is a local probe, the $\mu$SR signals from spatially different
areas of the sample are not averaged but superimposed in the measured spectra.
This feature allows to distinguish between the three above mentioned scenarios.
As discussed above, the ZF-$\mu$SR response of the magnetic areas of the sample
is characterized by a fast relaxing signal visible at early times of the
spectra, while a non-magnetic volume shows slow relaxation, better visible at
longer times, only.
In Fig.~\ref{fig:ZF-history}, two ZF muon-time spectra taken at the same
temperature ($T=2.6$~K) and pressure ($p=2.36$~GPa) with different magnetic
histories are shown.
The first muon-time spectra was recorded after cooling the sample
from $T\simeq100$~K to 2.6~K in zero magnetic field. By keeping
the temperature constant, the second ZF spectra was obtained after
ramping the magnetic field up to $\mu_0H\simeq0.1$~T and then
setting it back to zero.
Apparently, the ZF-$\mu$SR response of the magnetic areas of the sample, as
evidenced by the identically fast relaxations $\Lambda_{m,t1}$ and
$\Lambda_{m,t2}$ at early times of the spectrum (see Fig.~\ref{fig:ZF-history}b
and Table~\ref{tab1}), is not affected by the magnetic history.
On the contrary, the ZF-$\mu$SR signal representing the non-magnetic volume of
the sample exhibits a strongly larger relaxation $\Lambda_{\rm pm}$ after the
application of an external field at low temperatures (see
Fig.~\ref{fig:ZF-history}c and Table~\ref{tab1}).
%
%
%
This indicates that the superconductivity is most probably located within the
non-magnetic areas of the sample, since any changes of the magnetic field
within a superconductor with non-zero pinning leads to trapping the magnetic
flux and, as a consequence, to  a very  nonuniform field distribution inside
the superconducting parts of the sample \cite{Luke93}.
%

Our results point to a strong difference between LaFeAsO$_{1-x}$F$_x$ and the
structurally related families of Fe-based HTS with the La atom substituted by
other rare earths elements like Sm, Ce, Pr, Nd etc. In these families bulk
magnetism and bulk superconductivity are found  to coexist on the nanoscale
level \cite{Drew09,Sanna09,Sanna10}.
The magnetization and $\mu$SR experiments reveal that in LaFeAsO$_{1-x}$F$_x$,
the magnetism and superconductivity are not coexisting over the whole sample
volume, i.e. this system is inclined towards phase separation. The reduction of
the magnetic interaction and the simultaneous appearance of superconductivity
indicate a much stronger competition of the two ordered parameters.

\begin{table}[]
\begin{tabular}{c|ccc}
    &$\Lambda_{m,t1}$&$\Lambda_{m,t2}$&$\Lambda_{\rm pm}$\\
    &($\mu$s$^{-1}$)&($\mu$s$^{-1}$)&($\mu$s$^{-1}$)\\
\hline
ZFC                                 & 19.7(2.7)&1.28(14)&0.165(26) \\
ZFC$\rightarrow$0.1~T$\rightarrow$ZF& 18.4(3.5)&1.20(28)&1.20(36) \\
\end{tabular}
\caption{Parameters as extracted from the fit of Eq.~(\ref{eq:zf-musr}) to the
muon-time spectra obtained after cooling the sample from $T\simeq100$~K to
2.6~K in zero magnetic field (ZFC) and after ramping the magnetic field up to
$\mu_0H\simeq0.1$~T and setting it back to zero
(ZFC$\rightarrow$0.1~T$\rightarrow$ZF). $A_{\rm S}$, $A_{\rm PC}$, $\zeta$ and
$\omega$ were assumed to be the same for both spectra.} \label{tab1}
\end{table}

%
%

In conclusion, the interplay between magnetism and superconductivity was
studied in LaFeAs$_{0.945}$F$_{0.055}$ by performing muon-spin rotation and
magnetization experiments as a function of pressure up to $p\simeq 2.4$~GPa. At
ambient pressure the sample is purely magnetic, but at the border to the
superconducting state of LaFeAsO$_{1-x}$F$_{x}$. The application of hydrostatic
pressure leads to a substantial decrease of $T_{\rm N}$ and reduction of the
magnetic phase volume and, at the same time, to a strong increase of $T_c$ and
the diamagnetic susceptibility.
Magnetic history dependent ZF-$\mu$SR measurements show that superconductivity
most probably develops in the areas of the sample that are non-magnetic down to
lowest temperatures.
This clearly shows that in LaFeAsO$_{1-x}$F$_{x}$ magnetism and
superconductivity are competing order parameters.

The work was performed at the S$\mu$S,  Paul Scherrer Institute (PSI,
Switzerland) at the GPD instrument. The work of MB is supported by the Swiss
National Foundation (SNF). GP and SS acknowledge the support of NMI3 Access
Programme. RDR acknowledges MIUR PRIN 2008XWLWF9. The work of NDZ and SK was
partly supported by the NCCR program MaNEP.


\begin{thebibliography}{99}
%
\bibitem{Niedermayer98} C.~Niedermayer {\it et al.},
Phys.~Rev.~Lett. {\bf 80}, 3843 (1998).
%
\bibitem{Coneri10} F.~Coneri {\it et al.},
Phys.~Rev.~B {\bf 81}, 104507 (2010).
%
\bibitem{Khasanov_08-OIE-pahse-diagram} R.~Khasanov {\it et al.},
Phys.~Rev.~Lett. {\bf 101}, 077001 (2008).
%
\bibitem{Drew09} A.J.~Drew {\it et al.},
Nature~Mater. {\bf 8}, 310(2009).
%
\bibitem{Sanna09} S.~Sanna {\it et al.},
Phys.~Rev.~B {\bf 80}, 052503 (2009).
%
\bibitem{Nandi10} S.~Nandi {\it et al.},
Phys.~Rev.~Lett. {\bf 104}, 057006 (2010).
%
\bibitem{Khasanov09} R.~Khasanov {\it et al.},
Phys.~Rev.~B {\bf 80}, 140511 (2009).
%
\bibitem{Park09} J.T.~Park {\it et al.},
Phys.~Rev.~Lett. {\bf 102}, 117006 (2009).
%
\bibitem{Luetkens09} H.~Luetkens {\it et al.},
Nature~Mater. {\bf 8}, 305 (2009).
%
\bibitem{Sanna10} S.~Sanna {\it et al.},
Phys.~Rev.~B {\bf 82}, 060508 (2010).
%
\bibitem{Zhigadlo08} N.D.~Zhigadlo {\it et al.},
J.~Phys.:~Condens.~Matter {\bf 20}, 342202 (2008).
%
\bibitem{EasyLab} http://www.easylab.co.uk/.
%
\bibitem{Maeter09} H.~Maeter {\it et al.},
Phys.~Rev.~B {\bf 80}, 094524 (2009).
%
\bibitem{Luetkens08} H.~Luetkens {\it et al.},
Phys.~Rev.~Lett. {\bf 101}, 097009 (2008).
%
\bibitem{Khasanov08} R.~Khasanov {\it et al.},
Phys.~Rev.~B {\bf 78}, 220510 (2008).
%
%
\bibitem{Khasanov09_Ba122} R.~Khasanov {\it et al.},
Phys.~Rev.~Lett. {\bf 102}, 187005 (2009).
%
\bibitem{Vorontsov09} A.B.~Vorontsov {\it et al.},
Phys.~Rev.~B {\bf 79}, 060508 (2009).
%
\bibitem{Cvetkovic09} V.~Cvetkovic and Z.~Tesanovic, Phys.~Rev.~B {\bf 80}, 024512 (2009).
%
\bibitem{Bendele10} M.~Bendele {\it et al.},
Phys.~Rev.~Lett. {\bf 104}, 087003 (2010).
%
\bibitem{Savici02} A.T.~Savici {\it et al.},
Phys.~Rev.~B {\bf 66}, 014524 (2002).
%
\bibitem{Russo07} P.L.~Russo {\it et al.},
Phys.~Rev.~B {\bf 75}, 054511 (2007).
%
\bibitem{Luke93} G.M.~Luke {\it et al.},
Phys.~Rev.~Lett. {\bf 71}, 1466 (1993).
%
\bibitem{Zhigadlo11} N.~Zhigadlo {\it et al.}, under preparation.
%
\end{thebibliography}
\end{document}